# Radio emission from SN 1181 hosting a white dwarf merger product

Takatoshi Ko,[1,2,3,*] Daichi Tsuna ,[1,4] Bunyo Hatsukade,[5,6,7] and Toshikazu Shigeyama[1,2]

[1]Research Center for the Early Universe (RESCEU), School of Science, The University of Tokyo, 7-3-1 Hongo, Bunkyo-ku, Tokyo 113-0033, Japan
[2]Department of Astronomy, School of Science, The University of Tokyo, 7-3-1 Hongo, Bunkyo-ku, Tokyo 113-0033, Japan
[3]Astrophysical Big Bang Laboratory, RIKEN, 2-1 Hirosawa, Wako, Saitama 351-0198, Japan
[4]TAPIR, Mailcode 350-17, California Institute of Technology, Pasadena, CA 91125, USA
[5]National Astronomical Observatory of Japan, 2-21-1 Osawa, Mitaka, Tokyo 181-8588, Japan
[6]Graduate Institute for Advanced Studies, SOKENDAI, 2-21-1 Osawa, Mitaka, Tokyo 181-8588, Japan
[7]Institute of Astronomy, Graduate School of Science, The University of Tokyo, 2-21-1 Osawa, Mitaka, Tokyo 181-0015, Japan
*E-mail: ko-takatoshi@g.ecc.u-tokyo.ac.jp

**Abstract**

The remnant of the historical supernova 1181 is claimed to be associated with a white dwarf merger remnant J005311. The supernova remnant (SNR) shock, and a termination shock expected to be formed by the intense wind of J005311, are potential sites for radio emission via synchrotron emission from shock-accelerated electrons. In this paper, we estimate the radio emission from these two shocks, and find the peak radio flux to be 0.1–10 mJy (at 0.01–1 GHz) in the outer SNR shock and 0.01–0.1 mJy (at 1–10 GHz) in the inner termination shock. We also search for radio emission from this source in the archival data of the Karl G. Jansky Very Large Array (VLA) Sky Survey at 3 GHz, the NRAO VLA Sky Survey at 1.4 GHz and the Canadian Galactic Plane Survey at 408 MHz, finding no significant detection. While targeted observations with higher sensitivity are desired, we particularly encourage those at higher frequency and angular resolution to probe the inner termination shock and its evolution.

## 1 Introduction

Mergers of binary white dwarfs (WDs) evolve in different paths depending on the compositions and masses of the two WDs. Such mergers are thought to result in Type Ia supernovae (SNe) (Webbink 1984; Iben & Tutukov 1984) and/or neutron stars via accretion-induced collapse (Saio & Nomoto 1985; Taam & van den Heuvel 1986), but some can leave behind massive rapidly rotating WDs (Pshirkov et al. 2020; Caiazzo et al. 2021). The evolutionary paths of such WD merger products have started to be explored in detail (Schwab et al. 2016; Zhang et al. 2019; Wu et al. 2023; Yao et al. 2023; Zhong et al. 2024).

In 2019, an infrared nebula IRAS 00500+6713, hosting a massive WD J005311 at the center, was discovered (Gvaramadze et al. 2019). Spectroscopic observations revealed an intense wind blowing from the WD, with velocity $v_{\rm wind} \sim 16000\,{\rm km\,s^{-1}}$ and mass-loss rate $\dot{M}_{\rm wind} \sim 10^{-6}\,M_\odot\,{\rm yr}^{-1}$ (Gvaramadze et al. 2019; Lykou et al. 2023), supporting that this WD is a merger product (see also Kashiyama et al. 2019). Interestingly, the expansion rate and extent of the nebula, as well as its position on the celestial sphere and distance of $d = 2.3$ kpc (Bailer-Jones et al. 2021), makes J005311 the prime candidate for the remnant of the historical SN 1181 (Ritter et al. 2021).

Furthermore, X-ray data of this object obtained by XMM-Newton and Chandra and analyzed by Oskinova et al. (2020) and Ko et al. (2023) revealed that there is a more intense X-ray source at the center (of intensity $\sim 10^{-13}\,{\rm erg\,cm^{-2}\,s^{-1}\,arcsec^{-2}}$) in addition to the diffuse emission (of $\sim 10^{-17}\,{\rm erg\,cm^{-2}\,s^{-1}\,arcsec^{-2}}$) likely powered by the merger ejecta. While the XMM observations did not resolve the compact inner source, Chandra observations with a better angular resolution found that the inner source has a finite extent (0.0087–0.018 pc).

Ko et al. (2023) constructed a theoretical model that comprehensibly explains the multiple layers observed in various wavelengths (see their schematic figure and figure 1). They have newly suggested that the inner X-ray emission is powered by a termination shock, formed by the interaction between the unshocked SN ejecta and the fast wind from the central WD. They also confirm that the outer diffuse region is powered by the interaction between the SN 1181 ejecta and the surrounding interstellar medium (ISM), like SN remnants (SNRs). Using their model, they have succeeded in explaining the multi-wavelength observation properties, if the wind started blowing recently within the past 30 years.

SNRs and young SNe within years of explosion emit synchrotron radiation in radio, as the collisionless shocks formed by the interaction can accelerate particles and amplify the ambient magnetic field (e.g., Weiler et al. 2002; Dubner & Giacani 2015; Chevalier & Fransson 2017). This is likely the case for SN 1181 as well, but radio emission from this source has yet to be reported. Thus existing and future radio observations may offer a strong test for the model.

In this work, we consider radio emission from both the outer and inner shocked regions, which are schematically shown in figure 1. The radio source of interest is the region swept by the outer forward shock (as in typical SNRs) at $\theta_{\rm out} \approx 2'$, and the inner region swept by the wind termination shock at $\theta_{\rm in} \approx 1''$–$2''$. We first search for such signals in archival radio data that had covered this object. We next model the expected radio signal, and find that the radio flux in these regions may be bright enough to be detectable with deeper targeted follow-up by current telescopes.

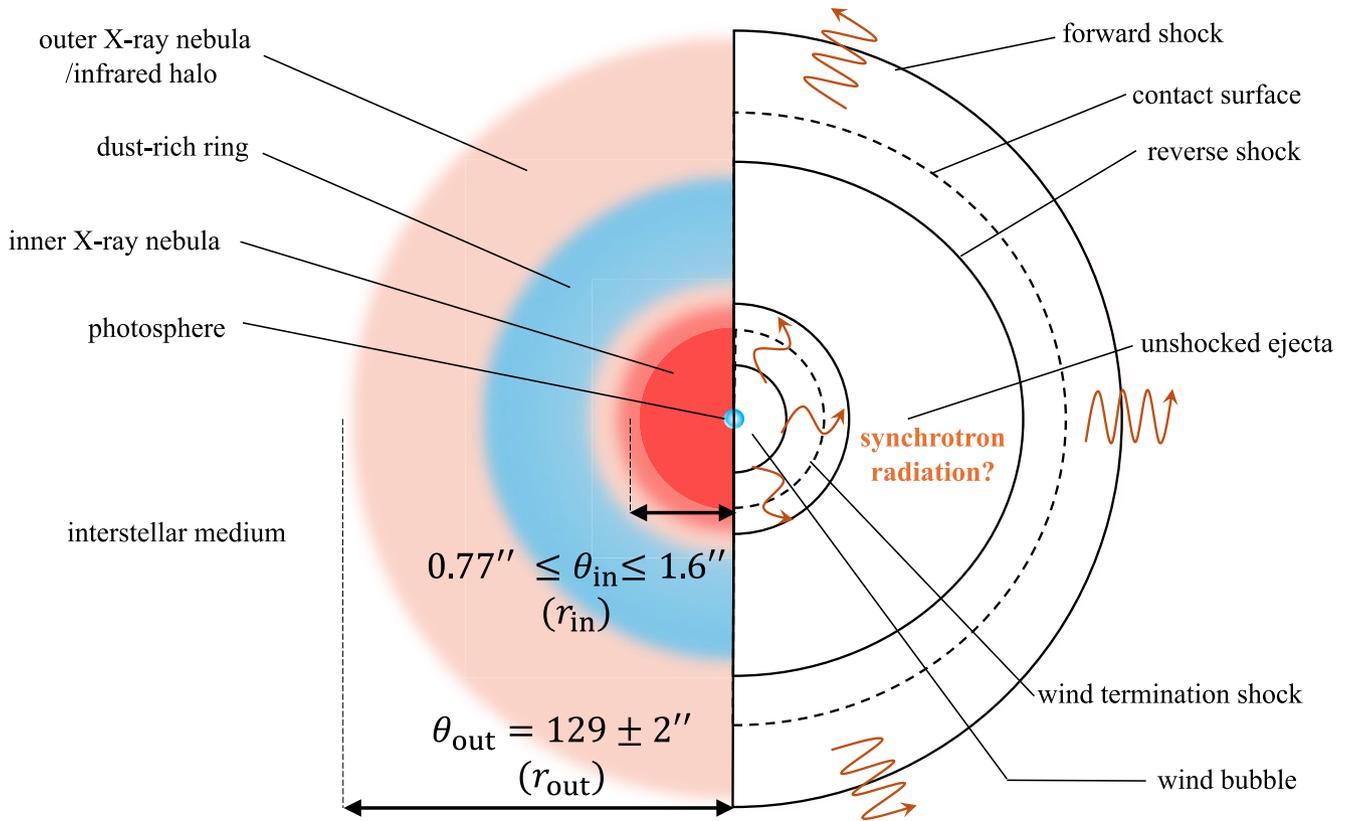

**Fig. 1.** Schematic picture of SNR 1181. The left-hand panel shows the observed features of this object, and the right-hand panel shows the explanations of each region based on the model proposed by Ko et al. (2023). We expect synchrotron radiation from the outer SN shock sweeping the interstellar medium (as in typical SNRs), and the WD wind termination shock sweeping the inner SN ejecta.

This work is structured as follows. In section 2 we present the archival radio observations and their results. Since no significant radio emission was found, we present upper limits to the radio flux for the inner and outer shocked regions. In section 3 we describe our model of the radio synchrotron emission. In section 4 we summarize our results, and mention future observability for the inner region, based on the time evolution of the emitting region and its radio luminosity.

## 2 Constraints from archival data

To constrain the radio emission, we search for radio observational data in the archive of the NRAO VLA Sky Survey (NVSS; Condon et al. 1998), the Very Large Array Sky Survey (VLASS; Lacy et al. 2020), and the Canadian Galactic Plane Survey (CGPS; Taylor et al. 2003; Tung et al. 2017). We specifically search for emission from the inner region with VLASS that has a smaller beam size of a few arcsec, while we mainly probe the outer region for the other two surveys with larger beam sizes.

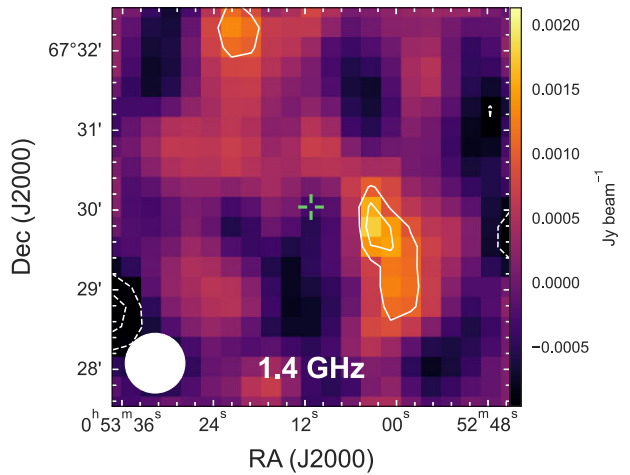

**Fig. 2.** NVSS 1.4 GHz continuum map centered at SN 1181 (cross mark). The contours are $\pm 2\sigma$ and $\pm 3\sigma$. The beam size is shown in the lower left-hand corner of the map.

NVSS is a 1.4 GHz continuum survey covering the entire sky north of $\sim 40°$ declination. The region containing SN 1181 was observed in 1993 November with a synthesized beam size of 45″ (full width at half maximum; FWHM). The rms noise level around the region is 0.48 mJy beam$^{-1}$. We find an emission with a peak flux density of 1.8 mJy at $\sim 50''$ east of the location of SN 1181 (figure 2). The source has a size of a few arcmin, most consistent with the outer shocked region (SNR) with a diameter of ∼240″ (Oskinova et al. 2020; Ko et al. 2023). Since the emission is marginal (signal-to-noise ratio of 3.7), we treat it as a nondetection and discuss it using the 3σ upper limit.

**Table 1.** Parameters relevant for the WD remnant, the X-ray nebula, and the radio emission modeling in this work.

| Parameters | Values |
|---|---|
| — Parameters from X-ray/optical (Gvaramadze et al. 2019; Ko et al. 2023) — | |
| Photosphere radius ($r_c$) | $1.0 \times 10^{10}$ cm |
| Photosphere temperature ($T_c$) | 211000 K |
| Inner shock radius ($r_\mathrm{in}$) | $5.6 \times 10^{16}$ cm |
| Outer shock radius ($r_\mathrm{out}$) | $4.5 \times 10^{18}$ cm |
| Upstream velocity in inner shock region ($v_\mathrm{in}$) | $1.5 \times 10^9$ cm s$^{-1}$ |
| Upstream velocity in outer shock region ($v_\mathrm{out}$) | $1 \times 10^8$ cm s$^{-1}$ |
| Mass loss rate from central WD ($\dot{M}_\mathrm{wind}$) | $5.0 \times 10^{19}$ g s$^{-1}$ |
| Density of outer shock region ($\rho_\mathrm{out}$) | $4.1 \times 10^{-25}$ g cm$^{-3}$ |
| Emission measure of inner shock region in X-ray (EM$_\mathrm{in}$) | $6.3 \times 10^{52}$ cm$^{-3}$ |
| Emission measure of outer shock region in X-ray (EM$_\mathrm{out}$) | $1.3 \times 10^{55}$ cm$^{-3}$ |
| Electron temperature of inner shock region ($k_B T_\mathrm{e,in,wind}$) | 1.2 keV |
| Electron temperature of outer shock region ($k_B T_\mathrm{e,out}$) | 0.3 keV |
| — Radio model parameters — | |
| Magnetic-field amplification efficiency ($\epsilon_B$) | 0.01 |
| Electron acceleration efficiency ($\epsilon_e$) | See table 2 |
| Power-law index of non-thermal electron ($p$) | 2.5 and 3 |

VLASS is the third radio survey project of the VLA designed to cover the sky visible at the VLA with a frequency bandwidth of 2–4 GHz. The region containing SN 1181 was observed three times: 2017 November, 2020 August, and 2023 March, with a synthesized beam size of 3″.1 × 2″.2, 3″.4 × 2″.2, and 3″.1 × 2″.4, respectively. The typical rms noise level of the Quick Look images is 0.14 mJy beam$^{-1}$ (Gordon et al. 2021). We do not find any significant emission. Since the largest angular scale structure of the S-band B-array observations is approximately 58″, which is smaller than the spatial extents of the outer shocked region (100″–120″), it is possible that the emission was resolved out in the observations.

CGPS observations at 408 MHz were conducted from 1995 to 2009 with the Synthesis Telescope at the Dominion Radio Astrophysical Observatory (Tung et al. 2017). The typical angular resolution is 2′.8 × 2′.8 cosec δ, and the typical sensitivity is 3 mJy beam$^{-1}$ rms. No significant emission is found.

## 3 Radio emission modeling

In a collisionless shock, we expect that magnetic fields are amplified and electrons are accelerated through diffusive shock acceleration. In this section we model the expected radio emission from the inner and outer shocked regions.

### 3.1 Synchrotron emission/absorption modeling

We model each shocked region as one-zone and parameterize the characteristic magnetic field strength $B_\mathrm{in}$ and $B_\mathrm{out}$ at each region, using an amplification efficiency $\epsilon_B$ as

$$\frac{B_i^2}{8\pi} = \epsilon_B \rho_i v_i^2. \qquad (1)$$

Hereafter $i = \{\mathrm{in, out}\}$ refers to quantities in the inner and outer shocked region, respectively. Here $\rho_i$ and $v_i$ are the upstream density and upstream velocity (in the shock rest frame) in region $i$, respectively. We fix the value of $\epsilon_B$ as $\epsilon_B = 0.01$, a typical value inferred for both SNe and SNRs (e.g., Chevalier & Fransson 2006; Maeda 2012; Murase et al. 2019; Reynolds et al. 2021).

The density of the WD wind powering the inner shocked region can be expressed as $\rho_\mathrm{wind} = \dot{M}_\mathrm{wind}/(4\pi r_\mathrm{in}^2 v_\mathrm{wind})$, where $\dot{M}_\mathrm{wind}$ and $v_\mathrm{wind}$ are, respectively, the mass-loss rate and velocity of the wind, and $r_\mathrm{in}$ is the radius of the inner shock front. For the outer shocked region, the relevant parameters $\rho_\mathrm{out}$ and $v_\mathrm{out}$ correspond to the ISM density and the SN forward shock velocity. From spectral analysis (Gvaramadze et al. 2019; Lykou et al. 2023) and dynamical modeling (Ko et al. 2023) of this object, the values of these parameters in 2021 A.D. were estimated to be those shown in table 1. Here, since the termination shock velocity (∼800 km s$^{-1}$ from our modeling) is much smaller than the wind velocity of $v_\mathrm{wind} \approx 1.5 \times 10^4$ km s$^{-1}$, we set $v_\mathrm{in} = v_\mathrm{wind}$. Here we adopt the fiducial case of the model; ($t_\mathrm{age,in}$, $M_\mathrm{ej}$) = (20 yr, 0.35 $M_\odot$).

Substituting values from table 1 into equation (1), the magnetic field strength at each region can be estimated as

$$B_\mathrm{in} \approx 0.70 \, \mathrm{mG} \left(\frac{\epsilon_B}{0.01}\right)^{1/2}, \quad B_\mathrm{out} \approx 32 \, \mu\mathrm{G} \left(\frac{\epsilon_B}{0.01}\right)^{1/2}. \qquad (2)$$

The maximum energy $E_\mathrm{max} = \gamma_\mathrm{max} m_e c^2$ of the electrons, where $\gamma_\mathrm{max}$ is the corresponding Lorentz factor, is set following Tsuna and Kawanaka (2019) by requiring the acceleration timescale (assuming the Bohm limit) $t_{\mathrm{acc},i} \approx (20/3)\gamma_i m_e c^3/(eB_i v_i^2)$ to be smaller than the age $t_{\mathrm{age},i}$ and the synchrotron cooling timescale $t_{\mathrm{cool},i} \approx 6\pi m_e c/(\sigma_T B_i^2 \gamma_i)$. Here $m_e$, $e$, $c$, and $\sigma_T$ are, respectively, the electron mass, the elementary charge, the speed of light, and the Thomson cross-section. A plausible range of $t_\mathrm{age,in} = 12$–30 yr is obtained from modeling of the inner shocked region (Ko et al. 2023), while $t_\mathrm{age,out} \approx 840$ yr is estimated from the SN date of 1181. The corresponding maximum frequencies of synchrotron radiation $\nu_\mathrm{max} = 3\gamma_\mathrm{max}^2 eB_i/16m_e c$ are $\nu_\mathrm{max,in} \sim 2 \times 10^{19}$ Hz and $\nu_\mathrm{max,out} \sim 7 \times 10^{16}$ Hz, which are much higher than the radio frequency of interest. Hence we expect that synchrotron emission spectra from accelerated electrons can easily be extended to the GHz band.

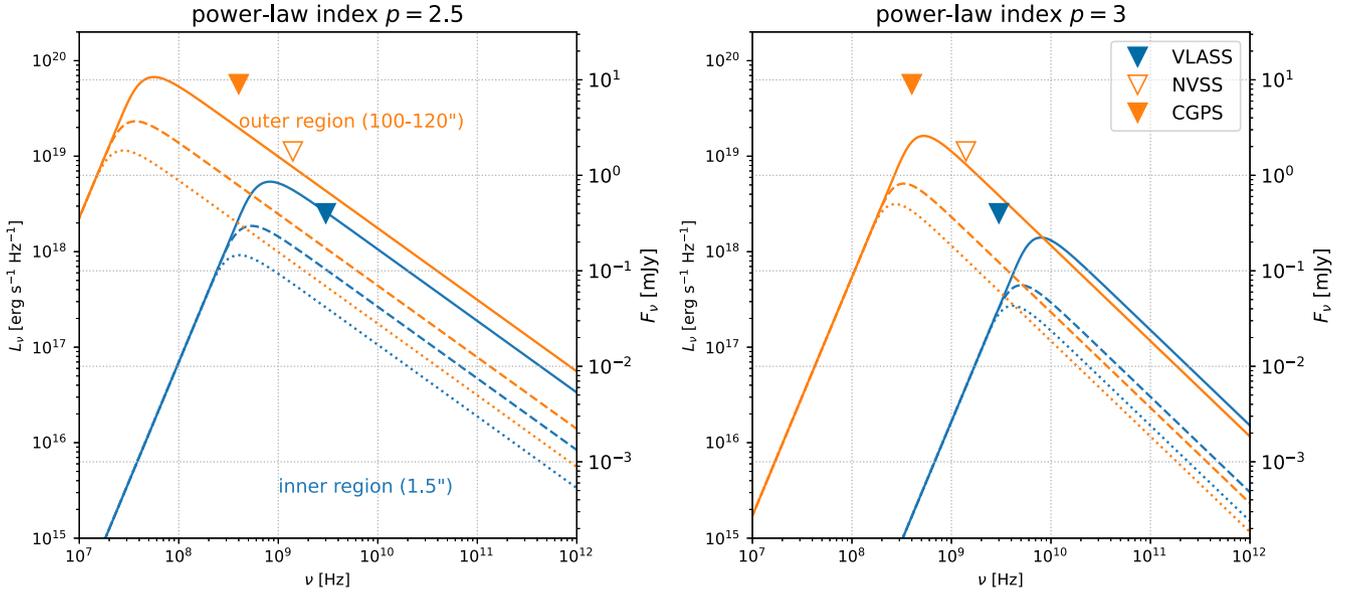

**Fig. 3.** Radio spectra for selected parameter sets in table 2. The blue and orange lines refer to the emissions from the inner and outer shocked regions, respectively. The solid, dashed, and dotted lines represent, respectively, the optimistic, fiducial, and pessimistic parameter sets. The triangles show the $3\sigma$ upper limits (with NVSS being a tentative detection), color-coded as constraints on the outer/inner region (for details see section 2). The horizontal gray line shows the $5\sigma$ sensitivity for Jansky VLA in the X-band (8–12 GHz), for an integration time of one hour. Note that the sensitivity applies to the total flux density when not spatially resolved, and a higher sensitivity is required when spatially resolved.

The relativistic electrons are assumed to follow a power-law energy distribution $N(E) \propto E^{-p}$ ($p > 2$), expected for diffusive shock acceleration. For an isotropic electron velocity distribution, the synchrotron power per frequency for each shocked region $i$ is (Rybicki & Lightman 1979):

$$P_{\text{syn},i}(\nu) = \Gamma\left(\frac{3p+19}{12}\right)\Gamma\left(\frac{3p-1}{12}\right)\Gamma\left(\frac{p+5}{4}\right)\left[\Gamma\left(\frac{p+7}{4}\right)\right]^{-1}$$
$$\times \frac{\sqrt{3}e^3 C_i B_i}{4\sqrt{\pi}m_e c^2(p+1)}\left(\frac{2\pi m_e c\nu}{3eB_i}\right)^{-(p-1)/2}, \quad (3)$$

where $\Gamma$ is the gamma function. Here $C_i$ is the normalization constant determined by the fraction of total dissipated energy given to relativistic electrons $\epsilon_e$ as

$$N_i(E) = (m_e c^2)^{p-1} C_i E^{-p} \quad (4)$$

$$\int_{E_{\min}}^{E_{\max}} E N_i(E) dE = \epsilon_e v_i^2 \int_0^{r_i} \rho_i(r) 4\pi r^2 dr, \quad (5)$$

where we assume $v_i$ is constant for simplicity and adopt the present-day value. We adopt a minimum electron energy $E_{\min} = 2m_e c^2$. For the power-law index $p$, while a range of $p = 2$–3 is adopted for an SNR in general, $p \approx 3$ is found from radio modeling for young SNRs (e.g., Chevalier & Fransson 2006; Maeda 2012). Here we consider $p = 2.5$ and $p = 3$.

It can be shown as follows that synchrotron cooling does not affect the radio emission. Finding the Lorentz factor $\gamma_{\text{cool},i}$, where $t_{\text{cool},i}$ is equal to the age, the frequencies $\nu_{c,i} = 3\gamma_{\text{cool},i}^2 eB_i/(16m_e c)$ beyond which cooling is important can be estimated for both regions as

$$\nu_{c,\text{in}} = 1.6 \times 10^{16} \text{ Hz} \left(\frac{\epsilon_B}{0.01}\right)^{-3/2}\left(\frac{t_{\text{age,in}}}{20 \text{ yr}}\right)^{-2}, \quad (6)$$

$$\nu_{c,\text{out}} = 9.5 \times 10^{16} \text{ Hz} \left(\frac{\epsilon_B}{0.01}\right)^{-3/2}\left(\frac{t_{\text{age,out}}}{840 \text{ yr}}\right)^{-2}. \quad (7)$$

At lower frequencies, the synchrotron emission is subject to synchrotron self-absorption. The absorption coefficient for a power-law distribution of electron energy [equation (5)] is given as (Rybicki & Lightman 1979):

$$\alpha_{\text{sa},i}(\nu) = \Gamma\left(\frac{3p+2}{12}\right)\Gamma\left(\frac{3p+22}{12}\right)\Gamma\left(\frac{p+6}{4}\right)\left[\Gamma\left(\frac{p+8}{4}\right)\right]^{-1}$$
$$\times \frac{\sqrt{3}e^3}{16\sqrt{\pi}m_e}\left(\frac{3e}{2\pi m_e^3 c^5}\right)^{p/2} \frac{C_i}{4\pi r_i^3/3} B_i^{(p+2)/2} \nu^{-(p+4)/2}, \quad (8)$$

and the optical depth is obtained as $\tau_{\text{sa},i}(\nu) = \alpha_{\text{sa},i}(\nu)r_i$.

Taking into account these absorption processes, the observed specific luminosity can be expressed as

$$L_{\nu,\text{in}} = \left(\frac{1 - e^{-\tau_{\text{sa,in}}}}{\tau_{\text{sa,in}}}\right) P_{\text{syn,in}}, \quad (9)$$

$$L_{\nu,\text{out}} = \left(\frac{1 - e^{-\tau_{\text{sa,out}}}}{\tau_{\text{sa,out}}}\right) P_{\text{syn,out}}. \quad (10)$$

The luminosities and fluxes (assuming the distance of 2.3 kpc) for sets of model parameters $(p, \epsilon_e)$ are shown in figure 3. The solid, dashed and dotted lines represent the optimistic, fiducial and pessimistic cases, respectively, which vary the values of $(p, \epsilon_e)$ as in table 2. The values for the fiducial cases are inspired from previous modeling of SNe and SNRs (e.g., Maeda 2012; Reynolds et al. 2021), and those for the optimistic cases are set to roughly explain the flux of the marginal finding from the NVSS image in section 2. The upper limits obtained from the previous surveys introduced in section 2 are shown as triangles. We find that for the outer shocked region the peak flux is about 0.1–10 mJy at frequencies around 0.1–1 GHz. For the inner shocked region, the peak flux is about 0.01–1 mJy around 1–10 GHz. Synchrotron self-absorption determines the rise at frequencies lower than the peak.

**Table 2.** Values of $p$, $\epsilon_e$ adopted in this work.

| $p = 2.5$ | $\epsilon_e$ | $p = 3$ | $\epsilon_e$ |
| --- | --- | --- | --- |
| Optimistic | $2 \times 10^{-3}$ | Optimistic | $5 \times 10^{-2}$ |
| Fiducial | $5 \times 10^{-4}$ | Fiducial | $1 \times 10^{-2}$ |
| Pessimistic | $2 \times 10^{-4}$ | Pessimistic | $5 \times 10^{-3}$ |

The flux becomes maximum when the optical depth is around unity. The frequency at the peak ($\nu_{\max,i}$) can be described as

$$\nu_{\max,\text{in}} \approx \begin{cases} 0.42 \text{ GHz} \left(\frac{\epsilon_B}{0.01}\right)^{-9/26} \left(\frac{\epsilon_e}{5 \times 10^{-4}}\right)^{-4/13} & (p = 2.5), \\ 4.3 \text{ GHz} \left(\frac{\epsilon_B}{0.01}\right)^{-5/14} \left(\frac{\epsilon_e}{0.01}\right)^{-2/7} & (p = 3), \end{cases} \quad (11)$$

and

$$\nu_{\max,\text{out}} \approx \begin{cases} 30 \text{ MHz} \left(\frac{\epsilon_B}{0.01}\right)^{-9/26} \left(\frac{\epsilon_e}{5 \times 10^{-4}}\right)^{-4/13} & (p = 2.5), \\ 0.30 \text{ GHz} \left(\frac{\epsilon_B}{0.01}\right)^{-5/14} \left(\frac{\epsilon_e}{0.01}\right)^{-2/7} & (p = 3). \end{cases} \quad (12)$$

Substituting them into equations (9) and (10), we obtain the maximum flux:

$$F_{\nu,\max,\text{in}} \approx \begin{cases} 3.4 \text{ mJy} \left(\frac{\epsilon_B}{0.01}\right)^{59/52} \left(\frac{\epsilon_e}{5 \times 10^{-4}}\right)^{16/13} & (p = 2.5), \\ 0.068 \text{ mJy} \left(\frac{\epsilon_B}{0.01}\right)^{19/14} \left(\frac{\epsilon_e}{0.01}\right)^{9/7} & (p = 3), \end{cases} \quad (13)$$

and

$$F_{\nu,\max,\text{out}} \approx \begin{cases} 0.27 \text{ mJy} \left(\frac{\epsilon_B}{0.01}\right)^{59/52} \left(\frac{\epsilon_e}{5 \times 10^{-4}}\right)^{16/13} & (p = 2.5), \\ 0.80 \text{ mJy} \left(\frac{\epsilon_B}{0.01}\right)^{19/14} \left(\frac{\epsilon_e}{0.01}\right)^{9/7} & (p = 3). \end{cases} \quad (14)$$

### 3.2 Free–free emission/absorption modeling

As the shocked regions contain charged particles due to shock heating/cooling, they may contribute to both free–free emission and absorption at radio frequencies. Here we infer free–free emission/absorption from three regions—the outer shocked region, the inner shocked WD wind and the inner shocked ejecta—and find that they are likely negligible in our situation.

For the shocked wind region, the flux of free–free emission is given (in cgs units) as (Rybicki & Lightman 1979):

$$F_{\nu,\text{in,wind}}^{\text{ff}} = 6.8 \times 10^{-38} T_{e,\text{in,wind}} e^{-h\nu/k_B T_{e,\text{in,wind}}}$$
$$\times \frac{1}{4\pi d^2} \int dV \sum_{\text{ion}} \left(Z_{\text{ion,in}}^2 n_{e,\text{in,wind}} n_{\text{ion,in,wind}}\right). \quad (15)$$

where $n_{e,\text{in,wind}}$ is the electron number density and $Z_{\text{ion,in}}$ and $n_{\text{ion,in,wind}}$ are the charge and number density of each ion, and $T_{e,\text{in,wind}}$ is the electron temperature of the inner shocked wind. As the inner shocked wind is responsible for the inner X-ray emission (Ko et al. 2023), this expression can be simplified using the observed emission measure

$$\text{EM}_{\text{in}} = \int dV \sum_{\text{ion}} (n_{e,\text{in,wind}} n_{\text{ion,in,wind}}) \quad (16)$$

and electron temperature $T_{e,\text{in,wind}}$. Approximating $h\nu/k_B T_{e,\text{in,wind}} \ll 1$ as valid in the radio band, where $h$ and $k_B$ denotes the Planck constant and the Boltzmann constant, respectively, we obtain

$$F_{\nu,\text{in,wind}}^{\text{ff}} \approx 6.8 \times 10^{-38} T_{e,\text{in,wind}} Z_{\text{ion,in}}^2 \frac{\text{EM}_{\text{in}}}{4\pi d^2} \quad (17)$$

$$\sim 0.77 \,\mu\text{Jy} \left(\frac{\text{EM}_{\text{in}}}{6.3 \times 10^{52} \text{ cm}^{-3}}\right) \left(\frac{k_B T_{e,\text{in,wind}}}{1.2 \text{ keV}}\right), \quad (18)$$

which is negligible compared to the synchrotron component in the frequency of interest. Here, we adopt $Z_{\text{ion,in}} = 8$, since the abundance of the inner shocked wind is oxygen-dominant (Ko et al. 2023). A similar calculation for the outer shocked region yields

$$F_{\nu,\text{out}}^{\text{ff}} \approx 6.8 \times 10^{-38} T_{e,\text{out}} Z_{\text{ion,out}}^2 \frac{\text{EM}_{\text{out}}}{4\pi d^2} \quad (19)$$

$$\sim 0.62 \,\mu\text{Jy} \left(\frac{\text{EM}_{\text{out}}}{1.3 \times 10^{55} \text{ cm}^{-3}}\right) \left(\frac{k_B T_{e,\text{out}}}{0.3 \text{ keV}}\right), \quad (20)$$

also negligible compared to the synchrotron component. Here, we adopt $Z_{\text{ion,out}} = 1$, since the abundance of the outer shocked region is near the solar abundance and hydrogen and helium are expected to be fully ionized (Ko et al. 2023).

The free–free absorption coefficient can be expressed as (Mezger & Henderson 1967)

$$\alpha_{\text{ff},i}(\nu) = 8.4 \times 10^{-28} \text{ cm}^{-1} \left(\frac{T_e}{10^4 \text{ K}}\right)^{-1.35} \left(\frac{\nu}{10 \text{ GHz}}\right)^{-2.1}$$
$$\times \sum \left(n_e/\text{cm}^{-3}\right) \left(n_{\text{ion}}/\text{cm}^{-3}\right) Z_{\text{ion}}^2, \quad (21)$$

where $n_e$ is the electron number density and $Z_{\text{ion}}$ and $n_{\text{ion}}$ are the charge and the number density of each ion. Evaluating the equation as was done for the free–free emission, we find that free–free absorption from these two regions are unimportant, with optical depths ($\sim \alpha_{\text{ff},i} r_i$) of

$$\tau_{\text{ff,in,wind}}(\nu) = 5.0 \times 10^{-12} \left(\frac{\nu}{10 \text{ GHz}}\right)^{-2.1}$$
$$\times \left(\frac{k_B T_{e,\text{in,wind}}}{1.2 \text{ keV}}\right)^{-1.35}, \quad (22)$$

$$\tau_{\text{ff,out}}(\nu) = 1.7 \times 10^{-14} \left(\frac{\nu}{10 \text{ GHz}}\right)^{-2.1} \left(\frac{k_B T_{e,\text{out}}}{0.3 \text{ keV}}\right)^{-1.35}, \quad (23)$$

both $\ll 1$ in the radio band.

The estimation for the inner shocked ejecta is less straightforward than the other two regions, as the electron temperature is expected to not exceed 1–a few eV, and thus its cooling emission would not contribute to the observed X-rays (see appendix A of Ko et al. 2023). The cooling timescale in this region is 3–4 orders of magnitude shorter than the age, and the region is expected to further compress into a thin shell by radiative cooling. This runaway cooling should continue until the shocked region recombines, and we may thus expect the free–free emission/absorption in this region to be negligible for an extremely small ionization fraction of $\sim 10^{-4}$–$10^{-3}$.

So far we have neglected the possibility that the shocked ejecta may become photo-ionized by the central WD that has a high effective temperature of $\sim 2\times 10^5$ K (Gvaramadze et al. 2019; Lykou et al. 2023). The optically thick wind from the WD may reprocess the ionizing photons from the central star and strongly suppress its flux, as found in models of optically thick winds from Wolf–Rayet stars (e.g., Smith et al. 2002; Götberg et al. 2017; Sander & Vink 2020). However, the situation can be complicated by the possible clumpiness of the wind and the ejecta, as well as the uncertainties in their composition and geometry.

Here we adopt a more empirical approach, and use the non-detection from the VLASS survey in section 2 to infer the ionization of the shocked ejecta. The radius, mass and density of the shocked ejecta are obtained by one-zone modeling (Ko et al. 2023) of a WD wind, with mass-loss rate of $\dot{M}_\mathrm{w}$, sweeping the bound part of the SN ejecta with a total mass of $M_\mathrm{ej}$ and following a density profile $\rho \propto r^{-1.5}$ at the innermost part (Tsuna et al. 2021). We adopt the model parameters in Ko et al. (2023) of $(\dot{M}_\mathrm{w}, t_\mathrm{age,in}, M_\mathrm{ej}) = (8\times 10^{-7}\,M_\odot\,\mathrm{yr}^{-1}, 20\,\mathrm{yr}, 0.35\,M_\odot)$ that reproduce the observed parameters in table 1, although the discussions below are not sensitive to the assumed values. By solving for the pressure in the thin shell at 2021 A.D. following subsection 3.3 of Ko et al. (2023), we find the corresponding gas density to be

$$\rho_\mathrm{in,ej} = \frac{\mu_\mathrm{ion} p_\mathrm{in,ej} m_u}{k_\mathrm{B} T_\mathrm{e,in,ej}}$$
$$\approx 2.4\times 10^{-16}\,\mathrm{g\,cm^{-3}}\left(\frac{\mu_\mathrm{ion}}{16}\right)\left(\frac{k_\mathrm{B} T_\mathrm{e,in,ej}}{1\,\mathrm{eV}}\right)^{-1}, \quad (24)$$

where $m_u$ is the atomic mass, $T_\mathrm{e,in,ej}$ is the electron temperature of the inner shocked ejecta, and $\mu_\mathrm{ion}$ is the mean molecular weight. We assume an oxygen-dominated composition for simplicity as in the Appendix of Ko et al. (2023), and that the ionized region is composed of neutral and singly-ionized oxygen with $\mu_\mathrm{ion} = 16(1 - 0.5 x_\mathrm{OII})$, where $x_\mathrm{OII}$ is the ionization fraction. The corresponding number density of electrons and oxygen ions is

$$n_\mathrm{e,ej} = n_\mathrm{ion,ej} = x_\mathrm{OII}\frac{\rho_\mathrm{in,ej}}{16 m_u}$$
$$\approx 9.0\times 10^6\,\mathrm{cm^{-3}} x_\mathrm{OII}\left(\frac{\mu_\mathrm{ion}}{16}\right)\left(\frac{k_\mathrm{B} T_\mathrm{e,in,ej}}{1\,\mathrm{eV}}\right)^{-1}. \quad (25)$$

The flux of free–free emission from the inner shocked ejecta is thus

$$F^\mathrm{ff}_{\nu,\mathrm{in,ej}} = 6.8\times 10^{-38} T_\mathrm{e,in,ej} e^{-h\nu/k_\mathrm{B} T_\mathrm{e,in,ej}}$$
$$\times \frac{1}{4\pi d^2}\int dV \sum_\mathrm{ion} n_\mathrm{e,in,wind} n_\mathrm{ion,in,wind}\Big)$$

$$\sim 200\,\mathrm{mJy}\,\left(\frac{x_\mathrm{OII}\mu_\mathrm{ion}}{8}\right)^2 \left(\frac{k_\mathrm{B} T_\mathrm{e,in,ej}}{1\,\mathrm{eV}}\right)^{-1}. \quad (26)$$

Here, we have adopted the inner shocked ejecta volume of

$$V_\mathrm{in,ej} = \frac{M_\mathrm{in,ej}}{\rho_\mathrm{in,ej}} \sim 1.8\times 10^{46}\,\mathrm{cm^3}, \quad (27)$$

where $M_\mathrm{in,ej}$ is the mass of the inner shocked ejecta.

The archive VLASS data constrains the emission to be less than 0.35 mJy, which requires the shocked ejecta to be almost neutral with $x_\mathrm{OII} \lesssim 2\times 10^{-2}$, consistent with the above expectation.[1]

However, the above scaling also shows that the free–free flux is sensitive to the value of $x_\mathrm{OII}$. Here we aim to obtain a conservative estimate for the radio detectability, and assume that free–free emission is negligible as expected for a plausible range of $x_\mathrm{OII} = 10^{-4}$–$10^{-3}$. On the other hand, the free–free absorption optical depth is given as

$$\alpha_{\mathrm{ff},i}(\nu)\frac{V_\mathrm{in,ej}}{4\pi r_\mathrm{in}^2} \sim 7\times 10^{-3}\left(\frac{\nu}{10\,\mathrm{GHz}}\right)^{-2.1}$$
$$\times \left(\frac{x_\mathrm{OII}\mu_\mathrm{ion}}{8}\right)^2 \left(\frac{k_\mathrm{B} T_\mathrm{e,in,ej}}{1\,\mathrm{eV}}\right)^{-3.35}, \quad (28)$$

and is thus expected to be negligible regardless of the exact value of $x_\mathrm{OII}$.

To summarize, we consider the free–free emission/absorption to be negligible in our radio modeling for all the shocked regions. While detailed photoionization calculations would be desired for better estimates of the free–free emission from the inner shocked region, future radio observations in multiple frequency bands may also be able to probe the relative contribution, as the spectral shape of free–free emission is different from that of synchrotron emission.

## 4 Discussion

We modeled radio signals from the shocked regions of the SN 1181 remnant, assuming that they are powered by synchrotron emission from electrons accelerated by the shocks (see also figure 1). Our search using archival radio data resulted in no significant signal, although one targeting the outer region may be a marginal detection.

Nevertheless, the radio signals we estimated may be bright enough to be detectable by targeted observations with current and future radio facilities. In particular, the possibility of detecting radio emissions from the inner shocked region is discussed here, since clarifying the radio emissions from the inner termination shock can constrain the nature of the fast-blowing wind. The inner region is generally bright at $\approx 10$ GHz regardless of the electron spectral index $p$, and for $p=3$ the peak lies around 10 GHz.

In this section, we estimate the future time evolution of the radio emission from the termination shock and discuss the observation strategy.

---

[1] Another possibility for the small radio flux is to have strong free–free absorption that suppresses the radio flux. However, for the free–free optical depth to exceed unity, oxygen should be in a high degree of ionization [equation (28)], and we disfavor this possibility as this should accompany strong and narrow ($\sim 800$ km s$^{-1}$) oxygen recombination lines that are not observed.

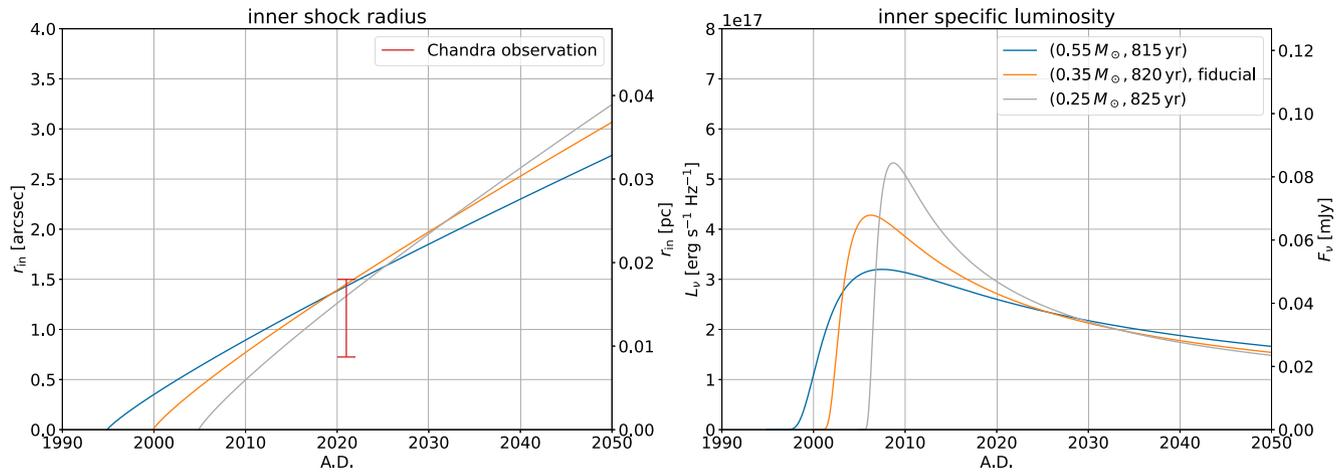

**Fig. 4.** Time evolution of the inner shock radius (left) and luminosity at 10 GHz (right). We fix $\dot{M}_{\rm wind}$ and $v_{\rm in}$ as values in table 1, and adopt the fiducial value of $\epsilon_e$ for $p = 2.5$. The red range in the left-hand panel shows the constraint of the shock radius obtained by Chandra in 2021 A.D.

### 4.1 Future observation prospects

We calculate the time evolution of the inner shocked region, and discuss prospects for future observations. We fix $\dot{M}_{\rm w}$ and adopt several parameter sets of $(t_{\rm w}, M_{\rm ej})$ which satisfy the constraints reported by Ko et al. (2023), including that for $r_{\rm in}$ obtained by high-resolution Chandra observations. Then we calculate the time evolution of $r_{\rm in}$ by solving the equations governing the shock dynamics in Ko et al. (2023), and also calculate the time evolution of $L_{\nu,\rm in}$ at 10 GHz, adopting the fiducial value of $\epsilon_e$ for $p = 2.5$ as an example. Here we have also neglected the free–free emission/absorption.

These results are shown in figure 4. The left-hand panel shows the time evolution of the radius for the parameter set $(t_{\rm age,in}, M_{\rm ej}) = (15\,{\rm yr},\ 0.25\,M_\odot), (20\,{\rm yr},\ 0.35\,M_\odot)$ which correspond to the models we adopted for figure 3, and $(25\,{\rm yr},\ 0.55\,M_\odot)$. This figure suggests that the source can be spatially resolved with a spatial resolution of at least $1''.5$.

For the light curve at 10 GHz in the right-hand panel, we generally find that the radio emission is gradually decaying at present. We also see that the earlier the wind was launched, the longer the rise time. This is because, for a nearly fixed $r_{\rm in}$, the earlier the wind launches, the slower the expansion rate, and the slower the rate of change of the absorption optical depth that governs the rise.

For a fiducial value of $\epsilon_e = 5 \times 10^{-4}$, the 10 GHz flux in the 2020s will typically be $\approx 30\,\mu$Jy, and will very weakly decrease over time in the decades that follow. This is the sensitivity required for current and future radio observations to detect the central radio emission.

For example, the Jansky VLA satisfies the required spatial resolution, so targeted observations may spatially resolve the inner shocked region for a sufficiently long exposure. In particular, next-generation telescopes with even higher sensitivity and angular resolution, such as the Square Kilometer Array[2] and the Next Generation VLA,[3] will achieve these requirements. These projects allow detailed studies of spatially resolved structures of the termination shock region, as well as long-term monitoring to probe the activity of the remnant WD over timescales as long as decades.

---

[2] ⟨https://www.skatelescope.org/⟩.
[3] ⟨https://ngvla.nrao.edu/⟩.

These high-resolution observations offer a unique probe for the formation mechanisms of the merger ejecta and the remnant wind. Throughout this paper, we assumed spherical symmetry, since the multiwavelength images of the inner and outer shocked regions are overall consistent with a spherical one. Nevertheless, numerical simulations of WD mergers aimed to reproduce SN Iax find the ejecta significantly departing from spherical symmetry (Kashyap et al. 2018, figure 4). Magnetohydrodynamical simulations modeling the central wind from a WD merger product also find asymmetry in the wind velocity profile (Zhong et al. 2024), which can lead to significant angular dependence in the radio power. If free–free emission from the shocked ejecta is important, multi-dimensional effects (e.g., mixing via Rayleigh–Taylor instabilities) can become important for characterizing the radio emission profile (D. Tsuna & X. Huang in preparation). With good spatial resolution (see figure 4), future observations may probe the asphericity of the innermost region and test these models.

### Acknowledgments

TK is grateful to Yuta Shiraishi and Kenta Hotokezaka for fruitful discussions. We thank Casey Law for guidance to the CGPS data. TK is supported by RIKEN Junior Research Associate Program. TS is supported by JSPS KAKENHI grant No. 22K03688, 22K03671, and 20H05639. DT is supported by the Sherman Fairchild Postdoctoral Fellowship at the California Institute of Technology. BH is supported by JSPS KAKENHI grant No. 19K03925 and 23K03449.

The National Radio Astronomy Observatory is a facility of the National Science Foundation operated under cooperative agreement by Associated Universities, Inc.